 \documentclass[%
reprint,
superscriptaddress,
 amsmath,amssymb,
pra,
floatfix
]{revtex4-2}
\usepackage{dsfont} 

\usepackage{graphicx}
\usepackage{bm}
\usepackage{color}
\usepackage{multirow}
\usepackage{textcomp}

\graphicspath{{./images/}{./}}

\definecolor{red}{rgb}{1,0,0}

\usepackage[colorlinks=true,linkcolor=blue,citecolor=blue]{hyperref}
\begin{document}

\preprint{APS/123-QED}
\title{Trypanosome motility in blood and particle suspensions}

\author{Florian A. Overberg}
\address{Theoretical Physics of Living Matter, Institute for Advanced Simulation, Forschungszentrum J\"ulich, 52425 J\"ulich, Germany}
\author{Marie Kater}
\address{Department of Cell and Developmental Biology, Biocenter, Julius-Maximilians-Universit\"at of W\"urzburg, W\"urzburg, Germany}
\author{Timothy Kr\"uger}
\address{Department of Cell and Developmental Biology, Biocenter, Julius-Maximilians-Universit\"at of W\"urzburg, W\"urzburg, Germany}
\author{Gerhard Gompper}
\address{Theoretical Physics of Living Matter, Institute for Advanced Simulation, Forschungszentrum J\"ulich, 52425 J\"ulich, Germany}
\author{Markus Engstler}
\address{Department of Cell and Developmental Biology, Biocenter, Julius-Maximilians-Universit\"at of W\"urzburg, W\"urzburg, Germany}
\author{Dmitry A. Fedosov}
\address{Theoretical Physics of Living Matter, Institute for Advanced Simulation, Forschungszentrum J\"ulich, 52425 J\"ulich, Germany}
 
\begin{abstract}
Motile microorganisms often navigate crowded and structurally complex environments, where surrounding obstacles can strongly influence locomotion. The bloodstream 
form of the flagellate parasite \textit{Trypanosoma brucei} circulates in blood, a dense suspension of red blood cells (RBCs), yet the physical mechanisms governing 
its locomotion under such conditions remain poorly understood because direct experimental observations are challenging. Here, we combine numerical simulations with 
in vitro experiments to investigate trypanosome motility in concentrated RBC suspensions and in suspensions of spherical colloidal particles. Simulations reveal that 
the parasite swimming speed increases by up to $50\%$ at RBC volume fractions comparable to those in blood. To identify the origin of this enhancement, we perform 
controlled studies in colloidal suspensions with particles of different sizes. We find that suspended particles substantially increase the anisotropy between 
the perpendicular and parallel friction coefficients acting on the beating flagellum, thereby enhancing propulsion. This effect is most pronounced when the suspended 
particles are comparable to or smaller than the characteristic wavelength of the flagellar beat. Experiments with microparticle suspensions confirm an increase in 
trypanosome propulsion with increasing particle concentration, in qualitative agreement with the simulations. Our results uncover a general physical mechanism by which 
concentrated particle suspensions can enhance flagellar-beat-driven locomotion and suggest that the densely crowded, particulate environment of blood may facilitate 
trypanosome propulsion. These findings provide new insight into trypanosome motility in the bloodstream and may apply broadly to other flagellated microswimmers in 
complex suspensions.
\end{abstract}

\maketitle

\section{Introduction}

Motile microorganisms, such as bacteria, sperm cells and parasites, must propel themselves through diverse and often complex environments, including viscoelastic mucus encountered 
by sperm cells \cite{Lauga_HSO_2009,Elgeti_PMS_2015} and intestinal contents traversed by some eukaryotic parasites \cite{Dawson_LEF_2010}. Many environments can impose challenging 
conditions, such as substantial crowding, complex material properties, spatial confinement, and dynamic flows. Microorganisms therefore require effective propulsion strategies that 
allow them to overcome these constraints and, in some cases, take advantage of the physical properties of their surroundings. For example, sperm cells can exhibit viscotaxis in viscosity 
gradients to navigate toward the egg \cite{Anand_VBF_2025} and the green algae \textit{Chlamydomonas reinhardtii} displays gravitactic behaviour that influences vertical distribution 
in the water column \cite{Kage_FGC_2020}.

A particularly interesting example is the flagellated eukaryotic parasite \textit{Trypanosoma brucei}, which undergoes major morphological and physiological adaptations as it alternates 
between its mammalian host and tsetse fly vector \cite{Schuster_DAT_2017,Schuster_UPT_2021,Szoor_TRY_2020}. In the mammalian host, proliferative bloodstream forms circulate in the blood 
and colonize extravascular tissues \cite{Bargul_SSA_2016,Krueger_BBT_2018}. Within the circulation, they encounter a highly crowded particulate environment in which red blood cells (RBCs) 
occupy approximately 40–50\% of the blood volume. Despite this crowding and the complex rheological properties of blood, bloodstream-form trypanosomes remain highly motile and can 
efficiently propel themselves through blood-like environments \cite{Bargul_SSA_2016,Broadhead_TRY_2006,Shimogawa_TRY_2018,Doro_VTV_2019}. 

Bloodstream-form \textit{T. brucei} cells have an elongated, asymmetric body and a single flagellum that emerges from the flagellar pocket near the posterior end of the cell, remains 
attached along most of the cell body through the flagellum attachment zone, and extends as a short free segment beyond the anterior end \cite{Hill_BMT_2003,Alizadehrad_SCD_2015}.
Propagation of flagellar waves along the attached flagellum deforms the elastic cell body and generates the characteristic rotational and helical swimming motion of the parasite 
\cite{Bargul_SSA_2016,Alizadehrad_SCD_2015,Overberg_TRY_2025}. During persistent forward swimming, the dominant flagellar wave propagates from the anterior tip toward the flagellar 
base, resulting in cell movement with the anterior end leading. Trypanosomes can also transiently reverse the direction of flagellar wave propagation and consequently move with 
the posterior end leading \cite{Baron_SIR_2007}. Such reversals can become particularly important in strongly confined or obstructed environments, where they allow cells to escape 
from obstacles or dead ends \cite{Heddergott_TMR_2012,Khameneh_TMR_2025}.

Numerous studies have investigated trypanosome locomotion in fluids of different viscosities examining changes in flagellar beating, swimming dynamics, cell body deformation, and 
directional reversals across different morphotypes and species 
\cite{Bargul_SSA_2016,Heddergott_TMR_2012,Wheeler_UCS_2017,Alizadehrad_SCD_2015,Overberg_TRY_2025,Schuster_DAT_2017,Khameneh_TMR_2025}.
These studies provide an important basis for understanding trypanosome locomotion in complex host environments. Direct experimental investigation of parasite motility in dense RBC 
suspensions, however, remains difficult because the high concentration and optical properties of blood cells strongly limit conventional microscopy. Additional challenges arise 
from the comparatively low abundance of parasites and the rapid dynamics of cells transported by blood flow.
 
A recent study reported the first live microscopy of \textit{Trypanosoma carassii} in the blood flow of zebrafish and showed that trypanosomes seem to be passively carried along 
by the flow of blood when the flow is strong, while for weak flows they can swim upstream \cite{Doro_VTV_2019}. In an earlier approach, trypanosome locomotion has been analysed in
microstructured environments containing arrays of obstacles designed to mimic key geometric aspects of a crowded bloodstream \cite{Heddergott_TMR_2012}. These experiments showed 
that the characteristic rotational motility of trypanosomes can promote efficient movement through densely obstructed environments, supporting the idea that trypanosome morphology 
and motility are adapted to the physical constraints encountered in the vertebrate host.
 
Numerical simulations are not subject to many of the experimental limitations associated with dense RBC suspensions and can provide detailed insight to the physical mechanisms governing
locomotion in complex suspensions. Recent simulations of \textit{T. brucei} under strong confinement, for example, have shown that moderate propulsion enhancement can occur when 
the confinement dimensions become comparable to the thickness of the parasite \cite{Tan_TMM_2025}. For a model swimming bacteria, enhanced swimming speeds have been reported 
in macromolecular polymer solutions \cite{Zoettl_EBS_2019}, where the effect arises from a non-uniform distribution of polymers close to the swimmer and an associated slip. 
Together, these findings illustrate that the physical structure of a complex environment can alter microswimmer propulsion in ways that cannot be explained by bulk viscosity alone.

In this study, we investigate trypanosome motility in RBC suspensions that reproduce key features of blood, and in controlled suspensions of spherical particles of different sizes. 
Both forward and backward motion is considered here. Even though backward motion is only transient in vivo, there are many examples of micro-organisms moving with either flagella or 
body first, so these mechanisms are of general interest. Simulations show that parasite propulsion speed significantly increases for both swimming directions in concentrated RBC 
suspensions comparable to physiological haematocrit. To identify the physical mechanism underlying the propulsion enhancement, we additionally performed forward motility simulations 
with spherical colloidal particles. We found that suspended particles strongly increase the anisotropy between the perpendicular and parallel friction experienced by the beating 
flagellum, thereby enhancing the propulsion. The effect is most pronounced when the suspended particles are comparable or smaller than the characteristic wavelength of the flagellar 
beat. The increase in trypanosome propulsion strength with increasing volume fraction of microparticles was also observed experimentally, in qualitative agreement with the simulation 
results. These results identify a general physical mechanism by which concentrated particulate environments can enhance flagellum-driven locomotion and provide a framework 
for understanding how the densely crowded environment of blood influences trypanosome motility.

\begin{figure}[ht]
    \centering
    \includegraphics[width=0.9\linewidth]{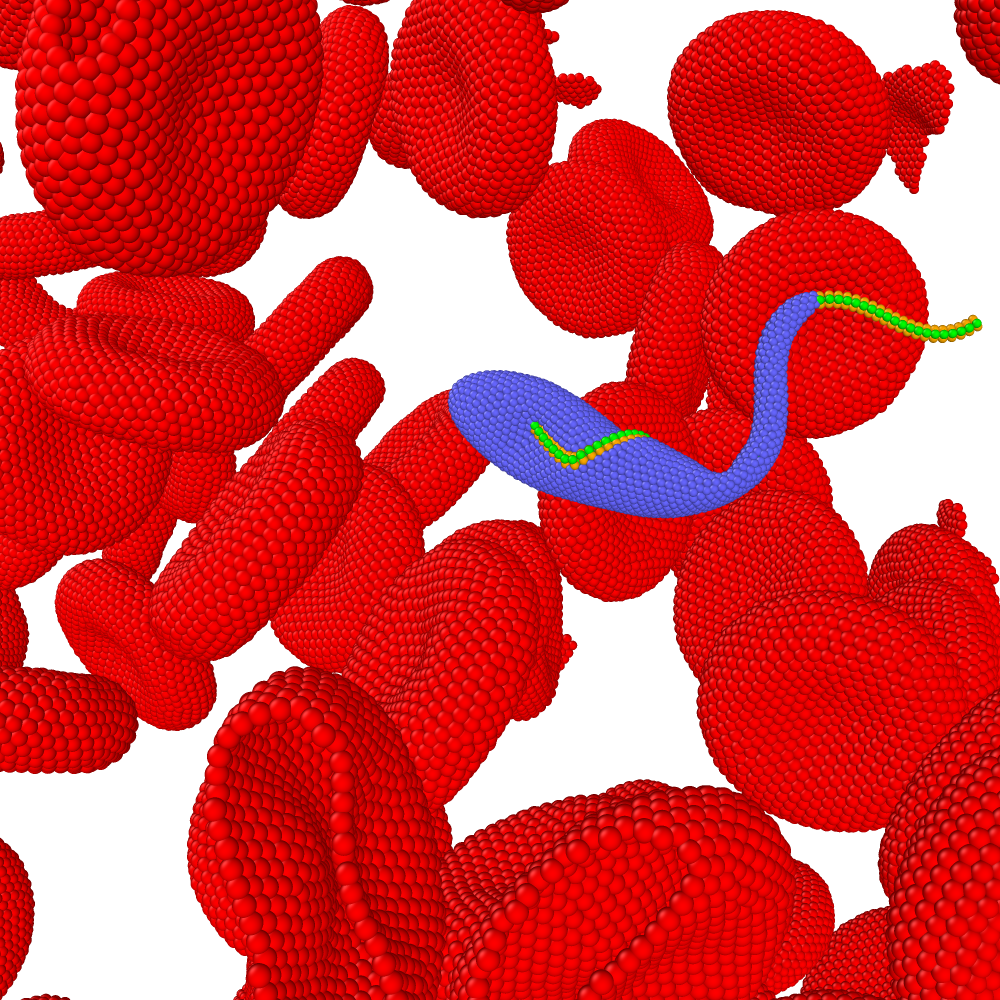}
    \caption{Trypanosome (with body in blue and flagellum in green and orange) within a suspension of RBCs (red) at $H_t = 0.1$ (see Movie S1). Some RBCs are 
    drawn as partially open surfaces, because they are located at the periodic boundaries.}
    \label{fig:tryp_RBC_susp_visualization}
\end{figure}

\section{Models \& methods}

\subsection{Models of RBCs and trypanosome}

RBCs are modeled as deformable particles, with a membrane enclosing a fluid volume \cite{Fedosov_RBC_2010, Fedosov_SCG_2010, Noguchi_STV_2005}. 
The membrane is represented by a triangulated network consisting of $N_v$ vertices, see Fig.~\ref{fig:tryp_RBC_susp_visualization}.
The vertices are interconnected by elastic springs that impose membrane shear elasticity. 
In addition to the in-plane elasticity, the RBC model 
incorporates bending elasticity and constraints for the conservation of surface area and volume. These constraints reflect key mechanical properties
of biological membranes: area incompressibility of the lipid bilayer and incompressibility of the enclosed cytoplasmic fluid. Together, these properties result in the cell's biconcave geometry.
A more detailed formulation of all energy terms for RBCs with corresponding parameters can be found in Appendix~\ref{app:rbc}.

The parasite model comprises a cell body with an attached flagellum, as shown in Fig.~\ref{fig:tryp_RBC_susp_visualization}. The trypanosome body
is modeled as an elastic membrane, which is similar to the model for RBCs (see Appendix~\ref{app:tryp}). The flagellum has a major part attached to 
the body, which is directly embedded into the membrane model of the body, and a short free portion at the anterior end of the parasite. The flagellum 
is constructed from four semiflexible filaments, whose dynamic modulation by extending and shortening springs within the flagellum structure leads to 
the emergence of a bending wave that provides parasite motility. For more details about the trypanosome model and it internal parameters, we refer to Appendix~\ref{app:tryp} and Ref.~\cite{Overberg_TRY_2025}. 

Excluded volume interactions between different RBCs are implemented through a repulsive part of the 12-6 Lennard-Jones (LJ) potential with parameters 
$\epsilon = 2.27\, k_BT$ and $\sigma_{LJ} =1.33 \times 10^{-2} L_{tryp}$. Here, $k_BT$ is a unit of thermal energy and $L_{tryp}$ is the trypanosome 
length at rest, which is employed as a length scale. For trypanosome-RBC interactions, $\sigma_{LJ} =2 \times 10^{-2} L_{tryp}$ to insure no penetration 
between RBCs and the parasite. Note that the distance cutoff of the LJ potential is such that it only includes repulsive interactions with no attraction.  

\subsection{Simulation setup}

The linear dimensions of the cubic simulation domain are $L_{domain} = 2L_{tryp}$ in all directions, with periodic boundary conditions. The volume fraction of RBCs (or hematocrit $H_t$)
is calculated as $H_t = N_{RBC} (V_{RBC} + \sigma_{LJ} A_{RBC}/2)/L_{domain}^3$, where $N_{RBC}$ is the number of RBCs, $V_{RBC}$ and $A_{RBC}$ are the RBC volume and 
surface area. Both cell types are embedded within a fluid represented by the smoothed dissipative particle dynamics (SDPD) method, a mesoscale hydrodynamics 
simulation technique derived through particle-based Lagrangian discretization of the Navier-Stokes equations \cite{Espanol_SDPD_2003,Mueller_SDPD_2015,Alizadehrad_SDPD_2018}. 
Fluid viscosity is set to $\eta = 7.67 \times 10^4 k_BT/(L_{tryp}^3 f)$, where $f$ is the beating frequency of the flagellum, which also defines a time scale. 
In physical units, fluid viscosity corresponds to blood plasma viscosity ($\eta = 1.2\, mPa\cdot s$) at physiological temperature \cite{Kesmarky_PVV_2008}. 
More details about the fluid model and its parameters can be found in Appendix~\ref{app:fluid}. 

For the trypanosome, the beating wave length $\lambda_{in}$ is set such that the number of simultaneously present waves $N_{waves} = L_{flag}/\lambda_{in} =1.75$, 
where $L_{flag}= 22.4 \, \mu m$ is the total flagellum length. Furthermore, the employed actuation amplitude results in the beating amplitude of $B_{0,sim}= 1.52\,\mu m$, which 
is calibrated by experimental measurements of the parasite beating characteristics at a viscosity of $1\, mPa\cdot s$ \cite{Khameneh_TMR_2025}. Further parasite
properties are listed in Table~\ref{table:trypanosome}.

Figure \ref{fig:tryp_RBC_susp_visualization} shows a simulation system at $H_t=0.1$, where the parasite, consisting  of body and attached flagellum,
is embedded in a suspension of RBCs (see Movie S1). Swimming speed $v$ of the trypanosome is measured by tracking the center of mass of the posterior 
cell body section and calculating the mean-squared displacement over time intervals $\Delta t_m=5\tau$ ($\tau = 1/f$). Rotation frequency $\Omega$ of the parasite is determined 
from the eigenvectors of the gyration tensor for the flagellar section after body wrapping. This approach allows us to identify the beating plane and its normal, 
from which orientation angles relative to a fixed reference vector can be computed. The resulting angles are constrained to the range $\theta \in [0, 2\pi]$, which  
accounts for a complete rotation. After that, $\Omega$ is calculated as the average of $\Delta \theta$ over $\Delta t_m$. The trypanosome 
is tracked for at least $80$ beats, during which it traverses multiple body lengths and passes through the periodic domain repeatedly, ensuring reliable averaging 
over independent configurations. 

In addition to a blood-like suspension of RBCs, trypanosome motility is studied within a suspension of hard colloidal spheres with different radii $R_{sph}$. 
Comparison is made for the same volume fraction of spheres and RBCs. In order to make spherical particles nearly non-deformable, the two-dimensional Young's modulus of the membrane is increased 
to $Y_{sph} = 10 Y_{RBC}$ and the target area is a few percent larger than $4\pi R_{sph}^2$, while the volume is a few percent smaller than $4/3\pi R_{sph}^3$, 
such that the membrane of a sphere is under significant tension. Deformation of spheres has been quantified to be below $1\%$, even when the parasite directly 
interacts with them. This is different for parasite-RBC interactions, where visible deformations of RBCs occur, consistently with experimental 
observations \cite{Bargul_SSA_2016}.

\subsection{Experimental setup for a sphere suspension}

Wildtype bloodstream form Trypanosoma brucei (Molteno Institute Trypanozoon antigenic type $1.6$) were cultivated in suspension at $37^oC$, $5\%$ $CO_2$ in 
Hirumi's modified iscove's medium 9, including a final 
volume of $10\%$ fetal calf serum (Sigma-Aldrich). Cells were kept in the exponential growth phase at a density less than $10^6\, cells/ml$. Polystyrene microbeads with 
diameters of $3\,\mu m$ (micromer-M, 08-02-303, Micromod) and $5\,\mu m$ (79633-F, Sigma-Aldrich) were used. Microscope chambers to hold $22\, mm\times 22 \,mm$ 
coverslips were drawn on slides using a pap pen. These were filled with $5\,\mu l$ of a 20x concentrated cell suspension mixed 1:1 with the respective 
microbeads and sealed with a coverslip. The resulting height of the chamber was $8\,\mu m \pm 0.5\, \mu m$, as determined by reference z-stacks recording 
the focus positions of the beads. Differential interference contrast imaging was performed using a Leica~DMI6000~B microscope, equipped with 
a 63x / NA 1.3 glycerol objective and a $37^oC$, $5\%$ $CO_2$ incubation chamber. Time series were recorded with a pco.edge sCMOS camera at 100 fps. 
Cells were recorded for maximally $45$ minutes per chamber.

Analysis of time series was performed using ImageJ. Regions of interest were selected that included the positions of a progressively forward swimming cell, 
exhibiting uninterrupted consecutive forward beats and translocating for at least one body length, as well as all beads in two bead diameters distance from 
the trajectory, which included all particles directly interacting with the trypanosome. The number of counted particles was multiplied by the calculated 
bead volume. This total bead volume was divided by the selected fluid volume to determine the volume fraction $\Phi$.

\begin{figure*}[ht]
    \centering
    \includegraphics[width=\textwidth]{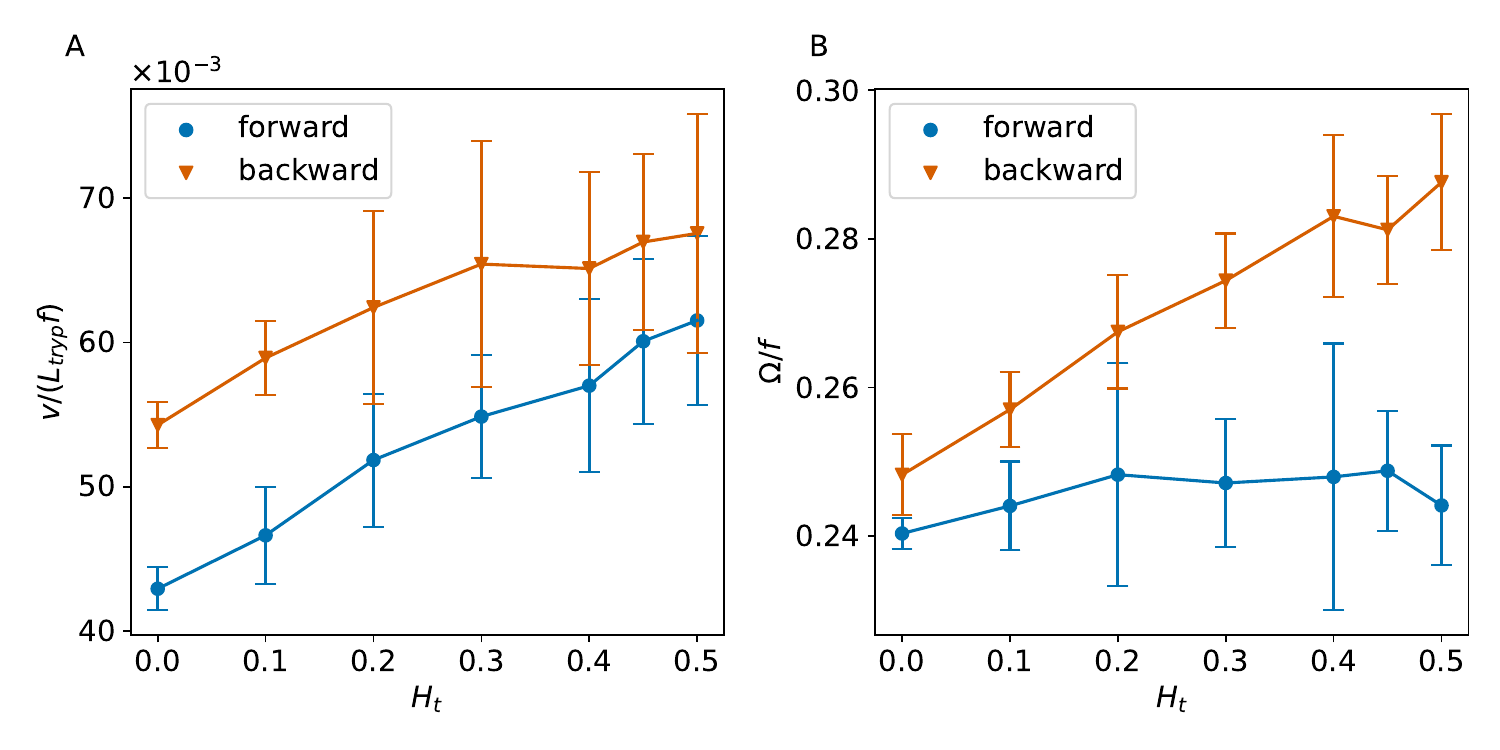}
    \caption{Swimming characteristics of the trypanosome in a RBC suspension as a function of hematocrit $H_t$. 
    A) Swimming velocity $v$ and 
    B) magnitude of the rotation 
    frequency $\Omega$. Circles (blue) represent forward motion and triangles (red) backward motion. The error bars correspond to the standard deviation of $v$ and $\Omega$.}
    \label{fig:RBC_susp}
\end{figure*}

\section{Results}

\subsection{Swimming speed in RBC suspensions}

We investigate trypanosome motility in RBC suspensions of various volume fractions for both forward and backward motion (see Movie S1). 
Figure~\ref{fig:RBC_susp}A shows that the swimming velocity increases with increasing $H_t$ for both forward and backward swimming. 
For forward motion, the swimming speed increases by about $43\%$ at $H_t=0.5$ in comparison to that in bulk fluid without RBCs. For backward motility, the corresponding 
increase in the swimming speed is roughly $24\%$. 
Our simulations show that at high enough $H_t$, the parasite leaves a cylinder-like track behind it, which forms due to the displacement of RBCs
by the parasite body. For forward swimming, the flagellum first interacts with dense RBCs in front, and then the body forces a cylinder-like track. 
In contrast, for backward swimming, the body forms a cylinder-like track first, followed by the beating flagellum within the track, which may reduce 
the interaction of the flagellum with surrounding RBCs, resulting in a poorer enhancement of the swimming velocity. 

Note that at $H_t=0$, the speed for backward motion is about $26 \%$ faster than for the forward 
motion. This is surprising, as the limited data available for persistent backward swimming shows slower speeds than for forward swimming trypanosomes \cite{Heddergott_TMR_2012}. 
There could be several differences between simulation and real cells that lead to this result. The model was developed to match parameters of persistently forward swimming 
cells \cite{Overberg_TRY_2025}, but the backward flagellar beat probably differs from a symmetrically reversed forward beat in several ways. 

It is remarkable that the parasite can swim faster with increasing hematocrit, even though RBC suspension becomes more viscous with increasing 
$H_t$. A qualitatively similar enhancement of the propulsion speed of trypanosomes has also been observed in polymer solutions with increasing polymer concentration 
\cite{Khameneh_TMR_2025}. These results indicate that the parasite can exploit surrounding structures for a faster propulsion.

The rotational behavior in Fig.~\ref{fig:RBC_susp}B differs markedly between forward and backward motion. Even though motion in both directions 
requires roughly $\sim 4$ beats per rotation in bulk fluid at $H_t=0$, the rotation frequency $\Omega$ for forward motion is nearly independent of 
hematocrit, while $\Omega$ increases with increasing $H_t$ for backward motility. Possible causes for this qualitative difference between forward 
and backward motion are discussed in detail in Sec.~\ref{sec:discussion}. Note that the rotation direction is reversed for opposing wave directions, and 
Fig.~\ref{fig:RBC_susp}B displays absolute values of $\Omega$. 

\begin{figure*}[ht]
    \centering
    \includegraphics[width=\textwidth]{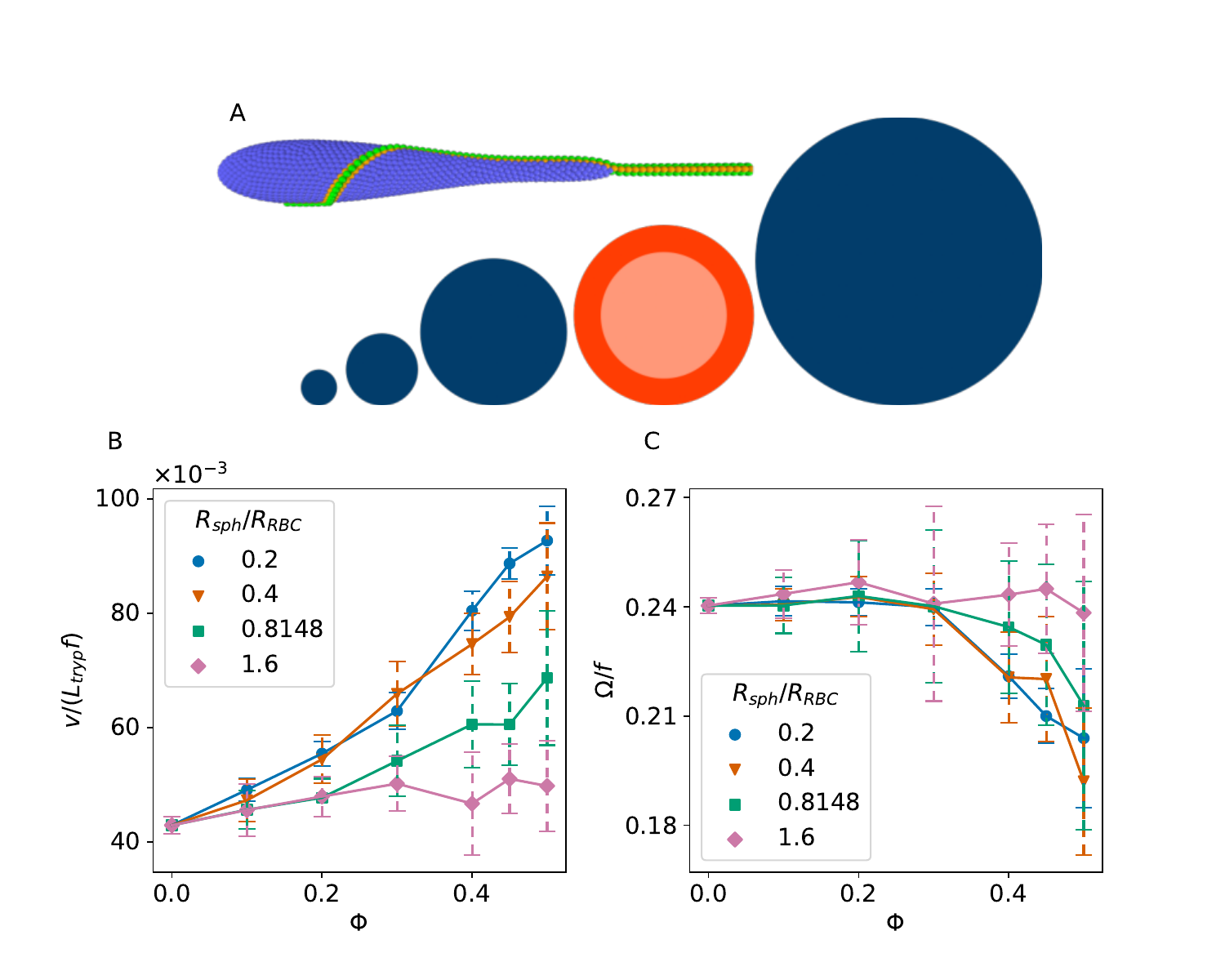}
    \caption{Trypanosome motility in colloidal suspensions (see Movies S2 and S3). A) Trypanosome with its unactuated shape in comparison to spheres (dark blue) of different 
    sizes and a RBC (red). B) Swimming velocity $v$ of the trypanosome in sphere suspensions as a function of volume fraction $\Phi$ for forward motion. 
    C) Rotation frequency $\Omega$ of the trypanosome in sphere suspensions as a function of $\Phi$ for forward motion. The error bars represent 
    the standard deviation of $v$ 
    and $\Omega$.}
    \label{fig:SPH_susp}
\end{figure*}
\begin{figure}[ht]
    \centering
    \includegraphics[width=0.95\linewidth]{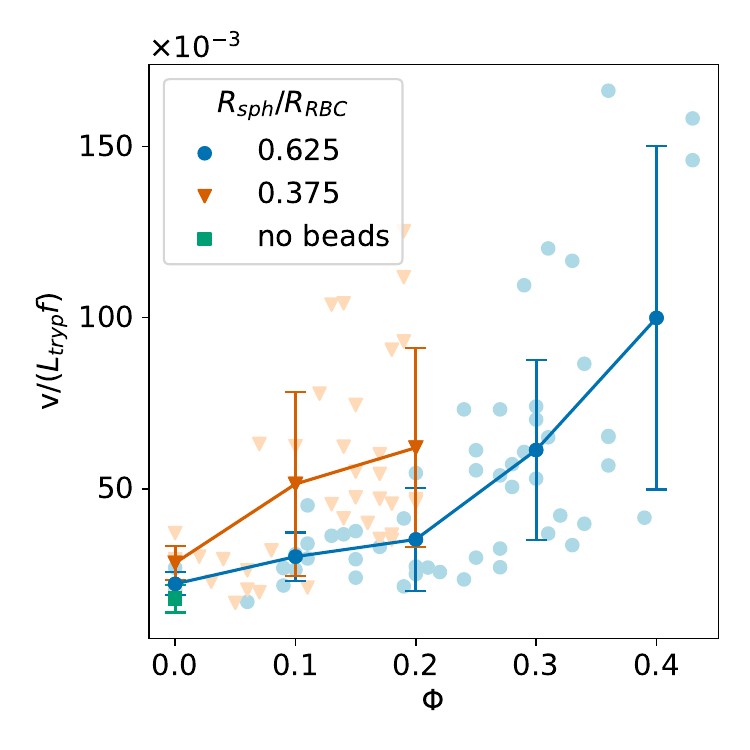}
    \caption{Experimentally measured swimming velocity of trypanosomes in a micro-bead suspension with varying volume fraction $\Phi$ for forward motion (see Movie S4). 
    The light blue symbols correspond to individual measurements in a suspension with $5\,\mu m$ diameter beads, while the dark blue symbols 
    represent the mean with standard deviation of the data, binned to the different $\Phi$ values. Both light orange and dark orange correspond to 
    results for suspensions with $3\,\mu m$ diameter beads. Additional measurements were made for a fluid without beads, for which the average 
    is plotted by green symbol.}
    \label{fig:SPH_susp_exp}
\end{figure}

\subsection{Swimming speed in colloidal suspensions}

To better understand the enhanced propulsion of the trypanosome in blood, we analyze its motion in suspensions of spherical colloidal particles of
different sizes (see Movies S2 and S3). Spherical particles are characterized by a single length scale, while RBCs have a more complex shape. Furthermore, RBCs are 
flexible, which may influence particle-trypanosome interactions, although the beating amplitude of the parasite remains largely unaffected 
by cell density in our simulations. Figure \ref{fig:SPH_susp}A compares spheres of various sizes and a RBC with the dimensions of the trypanosome. 
Here, the case of $R_{sph}/R_{RBC} = 0.815$ corresponds to $A_{RBC} = A_{sph}$ ($R_{RBC} = 4\,\mu m$ is the radius of a RBC). 
Figure \ref{fig:SPH_susp}B presents the corresponding forward swimming 
speeds of the trypanosome in sphere suspensions as a function of $\Phi$, with magnitudes comparable to those in RBC suspensions. For the largest 
sphere size $R_{sph}/R_{RBC} = 1.6$, the swimming speed shows no significant increase with increasing volume fraction. As the sphere size decreases, 
speed enhancement with increasing volume fraction becomes more pronounced. For the smallest sphere size $R_{sph}/R_{RBC} = 0.2$, the parasite 
speed at $\Phi=0.5$ is about $116 \%$ larger than that in bulk fluid without colloidal particles. The increase in parasite propulsion for 
the $R_{sph}/R_{RBC} = 0.2$ suspension is more than twice larger than that in RBC suspensions. 

Figure \ref{fig:SPH_susp}C shows parasite rotation frequencies as a function of $\Phi$ for different sphere sizes. For the largest spheres 
$R_{sph}/R_{RBC} = 1.6$, $\Omega$ is nearly independent of volume fraction, whose dependence is similar to that in RBC suspensions for forward 
motion of the trypanosome. For smaller sphere sizes, the rotation frequency $\Omega$ decreases with increasing volume fraction for large enough 
volume fractions $\Phi \gtrsim 0.3$.  

Experiments of trypanosomes swimming through a suspension of micro-beads of different diameters are 
performed to corroborate the simulations results (see Movie S4). Figure \ref{fig:SPH_susp_exp} shows an increase in the swimming velocity with increasing bead volume fraction, in agreement with 
the simulation results. Furthermore, the comparison of different bead sizes ($3\,\mu m$ and $5\,\mu m$) shows that an increase in the parasite 
propulsion is more pronounced for the small sphere size, again in qualitative agreement with the simulation results. The swimming speed $v$ 
at $\Phi=0$ is lower in experiments compared to simulations. However, an increase in the swimming speed at $\Phi=0.4$ in comparison to 
$\Phi=0$ is approximately four-fold in experiments, significantly more pronounced than the corresponding increase of $v$ in simulations as 
a function of $\Phi$. Note that experimental observations are performed in a slit-like setup with a height of $2-3$ bead sizes,
while simulations represent three-dimensional bulk systems. The slit-like setup in experiments is necessary to allow 
clean microscopic recordings and measurements, but introduces confinement, which differs from the bulk suspension considered in simulations, making a quantitative comparison difficult, 
though the qualitative agreement is very good. Interestingly, individual tracks in experiments show strong variability of
parasite speeds, with a pronounced increase when parasites enter a region with a high local packing fraction (see Movie S4). This speed increase at 
high enough local concentrations of suspended particles nicely illustrates their pronounced effect on parasite locomotion.  

\begin{figure*}[ht]
    \centering
    \includegraphics[width=\textwidth]{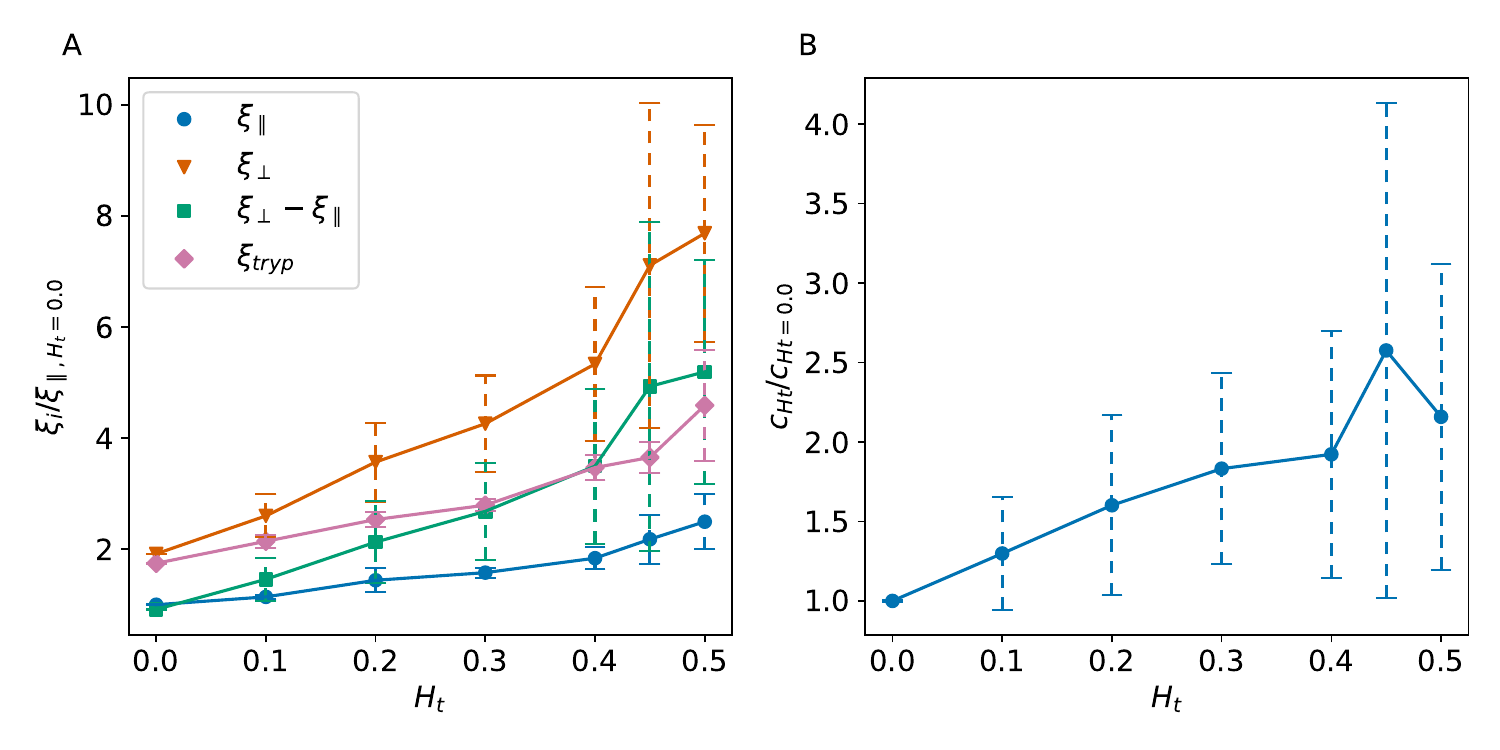}
    \caption{A) Friction coefficients in a RBC suspension as a function of hematocrit $H_t$, normalized by the parallel friction coefficient 
    $\xi_{\parallel, H_t=0.0}$ in a purely viscous fluid. 
    Symbols indicate friction coefficients $\xi_\parallel$ ($\circ$) and 
    $\xi_\perp$ ($\triangledown$) for a single passive flagellum,the difference $\xi_{\perp} - \xi_{\parallel}$ ($\square$), which is proportional to 
    the swimming speed, and the parallel friction coefficient $\xi_{tryp}$ ($\diamond$) for an unactuated trypanosome. 
    B) Coefficient 
    $c=(\xi_{\perp,H_t} - \xi_{\parallel,H_t} )/\xi_{tryp,H_t}$, which is proportional to propulsion strength 
    [see Eq.~(\ref{eq:vVsc})], as a function of $H_t$.}
    \label{fig:drag}
\end{figure*}

\subsection{Increased drag anisotropy as a main cause of propulsion enhancement}

A possible mechanism for an enhanced propulsion in suspensions with increasing volume fraction of particles is an increase in friction anisotropy. 
Flagellum propulsion force $F_{prop}$ directly depends on the difference of perpendicular ($\xi_\perp$) and parallel ($\xi_\parallel$) friction 
coefficients as $F_{prop}=(\xi_\perp - \xi_\parallel ) \cdot M$, where $M=\int_0^{L_{tryp}} \dot{y} \hat{y} dx$ 
is an integral over the flagellum 
length, $y$ is the flagellar displacement perpendicular to the swimming $x$-axis, and $\hat{y} = dy/dx$ is the local slope of the beat shape \cite{Lauga_HSO_2009, Elgeti_PMS_2015}. Assuming that the drag on 
the whole swimmer is given by $F_{drag} = \xi_{tryp}L_{tryp}v$ with the drag coefficient $\xi_{tryp}$, the swimming speed can be approximated as
\begin{equation}
\label{eq:vVsc}
    v = \frac{(\xi_{\perp} - \xi_{\parallel} )}{\xi_{tryp}}\cdot \frac{M}{L_{tryp}}.
\end{equation}
Since our simulations indicate that beating characteristics of the parasite are nearly independent of the volume fraction of suspended particles 
(e.g., a change in the beating amplitude remains within $2-3\%$ for increasing $\Phi$), 
it is plausible to expect that there must be significant changes in friction anisotropy (i.e., $\xi_{\perp} - \xi_{\parallel}$) with increasing 
$\Phi$ or $H_t$. 

To test this hypothesis, we perform simulations of dragging a straight (rod-like) passive flagellum of length $L_{flag}$ through RBC suspensions with different 
$H_t$. In these simulations, the flagellum remains stationary at a fixed position, while the RBC suspension is driven by a constant force $f_d$ on all 
particles and the time evolution of the global flow velocity $v_G$ is measured. The dynamics of the system can be described as 
\begin{equation}
    M\frac{\partial v_G}{\partial t} = A - Bv_G,
    \label{eq:drag_ODE}
\end{equation}
where $M=\rho V_{domain}$ is the total mass of the suspension, $A=f_d n_0 V_{domain}$ is the total driving force, and $B=\xi L_{flag}$ is 
the drag coefficient of the flagellum. The solution of Eq.~(\ref{eq:drag_ODE}) is given by
\begin{equation}
    v_G(t) = \frac{A}{B}-\frac{A}{B}e^{-Bt/M}.
    \label{eq:solDrag}
\end{equation}
For $H_t=0$, $\xi$ can simply be determined by taking $\lim_{t \to \infty} v_G$; however, for $H_t > 0$, pronounced fluctuations in $v_G$ are 
observed as the flagellum interacts with flowing RBCs. Therefore, for positive hematocrit values, the curve $v_G(t)$ is fitted over the whole 
simulation time to determine the drag coefficient $\xi$. The flagellum is placed perpendicular and parallel to the flow to measure perpendicular 
($\xi_\perp$) and parallel ($\xi_\parallel$) friction coefficients, respectively.  

Figure \ref{fig:drag}A shows the simulation results, where a significant increase in $\xi_\perp - \xi_\parallel$ is observed 
with increasing $H_t$. To make a better prediction for an increased propulsion strength with increasing $H_t$, we also measure the drag 
on the whole unactuated trypanosome $\xi_{tryp}$ using the same setup. $\xi_{tryp}$ at $H_t=0.5$ is approximately twice larger than in bulk 
fluid without RBCs. Then, Fig.~\ref{fig:drag}B presents the propulsion strength $c=(\xi_{\perp,H_t} - \xi_{\parallel,H_t} )/\xi_{tryp, H_t}$,
which increases by about a factor of two with increasing $H_t$, suggesting that $v_{H_t=0.5}/v_{H_t=0.0} \approx 2$. An enhancement of 
propulsion by a factor of two is larger than our simulation results for RBC suspensions. Clearly, the friction estimation for a straight (rod-like) 
passive flagellum is not expected to provide a 
quantitatively accurate replication of drag coefficients on a dynamically moving parasite. However, these measurements qualitatively support
that an increase in drag anisotropy with increasing $H_t$ is primarily responsible for the observed enhanced propulsion of the parasite.

\section{Discussion \& conclusions}
\label{sec:discussion}

In this study, we have shown both in simulations and experiments that trypanosomes can utilize suspended particles of several micron in size
to enhance propulsion and increase their swimming speed. This ability enables the parasites to efficiently traverse through dense suspensions 
of cells such as blood, even though exactly the opposite may have been expected due to an increase in bulk viscosity and crowding for large
volume fractions of cells. In addition, an enhanced swimming speed leads to increased shear stresses at the parasite surface, which facilitates 
the transport of VSG proteins toward the flagellar pocket, where their internalization aids in the evasion of immune system \cite{Engstler_TRY_2007}. 
Our results are also qualitatively consistent with experiments of trypanosome motility in pillar arrays \cite{Heddergott_TMR_2012,Bargul_SSA_2016}, 
which result in an increased swimming speed for certain distances between pillars. In these experiments, the fixed pillars were interpreted as 
a "frozen suspension" of blood cells.

To clarify the physical mechanism responsible for speed enhancement with increasing volume fraction, we turned to a suspension of hard spheres of different
sizes, which are non-deformable and are characterized by a single length scale in comparison to RBCs. A pronounced enhancement of parasite swimming speed 
with increasing sphere volume fraction has been observed for small enough colloid sizes ($R_{sph} \lesssim 4-5\,\mu m$). Note that this critical size 
is comparable with the half wave length ($\lambda /2\approx 4.75\,\mu m$) of the beating flagellum. This observation strongly suggests that suspended 
particles must "fit" into wave pockets of the beating flagellum to provide a significant enhancement of parasite motility. In fact, when the colloidal
particles increasingly fit within wave pockets, the parasite can use them to "push" itself forward, similar to the propulsion enhancement found in 
pillar arrays \cite{Heddergott_TMR_2012,Bargul_SSA_2016}. For the case of large spheres with a diameter of $12.8\, \mu m$, the parasite is not able 
to exploit them efficiently for propulsion. Recent simulations of trypanosome motility in tubes 
\cite{Tan_TMM_2025} show that an increase in the swimming speed is possible under strong confinements due to hydrodynamic interactions of the beating 
flagellum with solid boundaries. Note that for small spheres, the velocity enhancement saturates with increasing volume fraction $\Phi$, 
since wave pockets of the beating flagellum are fully filled with suspended particles, and no further improvement is possible. 
In the limit of $R_{sph} \to 0$, the surrounding fluid is expected to become homogeneous, such that there should not be a significant propulsion 
enhancement. As a result, the speed enhancement is expected to become weaker for suspensions with much smaller particles than those used in simulations
and experiments. It is also worth mentioning that at high enough volume fractions (e.g., approaching the close-packing limit for sphere suspensions), 
the swimming ability of trypanosomes should become significantly restricted, as they are expected to get stuck within a very crowded environment.  

Even though it might seem intuitive, the statement above that parasites can use suspended particles for propulsion needs further clarification. 
When a beating wave runs along the flagellum, suspended particles within the wave pockets are displaced by the running wave. Their displacement by 
the flagellum is counter-balanced by viscous drag forces, suggesting that these particles significantly increase the drag on the flagellum in 
the normal direction. Since the beating characteristics of the flagellum are nearly independent of particle volume fraction, it plausible to assume 
that an increase in the difference between normal and tangential friction coefficients of the flagellum (presumably due to an increase in the normal 
friction) is the main physical mechanism responsible for propulsion enhancement. For a single unactuated flagellum, we have verified that the normal 
friction coefficient increases substantially with increasing volume fraction of suspended particles, while the tangential friction coefficient shows 
only a moderate increase. This confirms the proposition that suspended particles serve as local friction points that strongly increase friction 
anisotropy of the flagellum and enhance trypanosome propulsion. Note that the friction anisotropy of an unactuated flagellum is significantly increased 
with increasing $\Phi$ for suspensions of small as well as large colloids. However, this mechanism can be used by a beating flagellum only when 
suspended particles are small enough and fit into the wave pockets locally.  

An interesting aspect is the difference in velocity between forward and backward swimming. Already for $H_t=0$, this difference exists, which seems 
counter-intuitive for low Reynolds number flow (i.e., time-reversible flow), since for a single flagellum, the velocity is independent of the swimming 
direction. Simulations of an unactuated trypanosome dragged in different directions have also shown that there is no directional dependence in friction 
coefficients. To better understand the reason for the directional difference in swimming speed, we have turned to a simplified swimmer constructed from 
a single straight flagellum with an attached spherical head. This model shows that mechanical interaction between the attached body and the 
flagellum affects the beating wave, depending on its propagation direction. Effectively, this interaction tilts the axis of bending wave propagation. 
When the beating wave moves away from the body (i.e., a pusher swimming with the body first), the wave axis is tilted more than when the beating wave
moves toward the body (i.e., a puller swimming with the flagellum first). A more tilted wave axis effectively results in an increased wave amplitude, 
leading to a larger swimming velocity of the pusher in comparison to the puller. Even though 
the attachment of the flagellum to the trypanosome body is more complex than in this simplified model, the corresponding effect must be similar. 
It is difficult to observe it directly for trypanosomes, since the body deforms due to flagellum actuation, not allowing to clearly capture 
differences in the effective beating amplitude. 

Our simulations show that forward swimming in blood is not beneficial in terms of propulsion velocity in comparison with backward swimming. Note that in our model,
only the direction of wave propagation is changed for forward and backward motion, while all other parameters are kept unchanged. For real trypanosomes,
backward and forward wave actuation dynamics (e.g, amplitude, frequency) may not be identical \cite{Branche_CSF_2006}, which would also affect the corresponding 
swimming velocities. Clearly, for both forward and backward motion, the parasite can use RBCs for propulsion enhancement, as shown in Fig.~\ref{fig:RBC_susp}A.   

Rotation frequency of the modeled parasite slightly exceeds the value measured experimentally in bulk fluid with a viscosity $\eta = 1\, mPas$ \cite{Khameneh_TMR_2025}. 
For small spherical particles, parasite rotation for forward motion is slowed down with increasing volume fraction, which is likely a consequence of crowding such that friction 
on the rotating beating plane is increased. 
For large spherical colloids and RBCs, parasite rotation is nearly independent of $\Phi$ or $H_t$. For large spheres,
the gaps between them are sufficiently large for the parasite to rotate without significant hindrance. For RBCs, their local deformation eases trypanosome 
rotation, so that the RBCs do not have to be significantly displaced. When the parasite swims backwards, its rotation is slightly elevated with increasing hematocrit. 
Analysis of parasite deformation during swimming shows that in this case, the trypanosome body is slightly more bent than for the forward motion, 
which is also the case in bulk fluid \cite{Overberg_TRY_2025}. This bending chirality of the parasite is primarily responsible for its rotation, and is consistent 
with a slight enhancement of the parasite rotation with increasing $H_t$.

It is also worth mentioning some limitations of our experiments, which restricts the accessible range of parameters. First of all, the experimental system
corresponds to a slit with a thickness of several sphere diameters, because observations of trypanosome motion within a dense suspension in much thicker channels are not feasible. This means that the 
parasites may often interact with solid boundaries of the slit. Nevertheless, the mechanism for propulsion enhancement is expected to 
be similar in this case.  Furthermore, experimental observations show
that suspended particles are not always homogeneously distributed within the channel, especially at lower concentrations. Nevertheless, different portions of recorded 
tracks corresponding to dense and less dense regions are analysed separately, with a local estimation of particle concentration.   

The choice of beating characteristics of the modeled trypanosome is motivated by experimental measurements of parasite motility in a fluid with a viscosity $1\, mPas$,
which is close to blood plasma viscosity. The beating of trypanosomes in blood or a suspension of spheres might have different wave characteristics (e.g., different 
wave length or amplitude). Furthermore, other trypanosome species have different cell geometry and beating properties. These differences will of course affect 
quantitative characteristics of trypanosome propulsion in blood or concentrated suspensions. However, the discussed mechanism for propulsion enhancement should 
remain generally applicable to different trypanosme species. More broadly, this mechanism must also be applicable to a much broader class of microorganisms, which 
propel using external appendages like actuated flagella. 

Finally, in mammalian infections, trypanosomes do not encounter stationary RBC suspensions as considered in this study, but instead swim in blood vessels under 
considerable flow \cite{Doro_VTV_2019}. Nevertheless, it is important to understand the effect of suspended particles such as RBCs on the swimming behavior of 
trypanosomes. Flowing blood presents an even more complex environment with additional effects, such as cross-stream migration, shear gradients, and rotational 
stresses. Therefore, it would be difficult to isolate different effects in a blood flow environment, and provide physical insights for trypanosome propulsion in 
a complex suspension. However, the investigation of trypanosome behavior in blood flow is one of the necessary next steps toward understanding parasite behavior 
within its natural host.

\section*{Author Contributions} 

G.G. and D.A.F. conceived the research project. F.A.O. performed the simulations and analysed the obtained data.  
M.K. and T.K. performed experiments and interpreted experimental data.
All authors participated in the discussions and writing of the manuscript.

\section*{Conflicts of interest}

The authors have declared that no competing interests exist.

\section*{Acknowledgements}

Support by the Deutsche Forschungsgemeinschaft (DFG) within the Priority Programme "Physics of Parasitism" (SPP 2332) is gratefully acknowledged.
The authors gratefully appreciate computing time on the supercomputer JURECA \cite{jureca} at Forschungszentrum J{\"u}lich under grant no. actsys.

\section*{Data availability}

The data that support the findings of this article are not publicly available. The data are available from the authors upon reasonable request.

\appendix

\section{Trypanosome model}
\label{app:tryp}

The parasite body is modeled as a closed elastic membrane, which is implemented as a triangulated surface whose vertices 
interact via a set of energy potentials \cite{Overberg_TRY_2025}. $N_v$ vertices in the triangulation are connected by $N_b$ harmonic
springs with a potential 
\begin{equation}
    U_{bond} = \frac{1}{2}k_{s}\sum_{i}^{N_b}(l^i - l_0^i)^2,
    \label{eq:bond}
\end{equation}
where $k_{s}$ denotes the spring stiffness, $l^i$ represents the spring length, and $l_0^i$ is the equilibrium bond length 
of spring $i$, set after initial triangulation to ensure a stress-free body at rest. The bending elasticity is implemented 
through a Helfrich bending energy potential \cite{Helfrich_EPB_1973} whose discretization is given by 
\begin{equation}
U_{bend} = \frac{\kappa}{2}\sum_i^{N_v} \sigma_i(H_i-H_0^i)^2,
\end{equation}
where $\kappa$ is the bending rigidity of the membrane, $\sigma_i=\sum_{j(i)} \sigma_{ij} r_{ij}/4$ is the area corresponding to 
vertex $i$ in the membrane triangulation, $H_i = \mathbf{n}_i\cdot\sum_{j(i)}\sigma_{ij}\mathbf{r}_{ij}/(\sigma_ir_{ij})$ is the mean 
curvature at vertex $i$, and $H_0^i$ is the spontaneous curvature at vertex $i$. Here, $\mathbf{r}_{ij} = \mathbf{r}_i - \mathbf{r}_j$,
$r_{ij}=|\mathbf{r}_{ij}|$ and $j(i)$ corresponds to all vertices linked to vertex $i$. $\sigma_{ij} = r_{ij}(\cot\theta_1 + \cot \theta_2)/2$ 
represents the bond length in the dual lattice, where $\theta_1$ and $\theta_2$ are the angles at the two vertices opposite to the edge 
$ij$ in the dihedral. 

In addition to the elasticity potentials above, volume and surface area conservation potentials \cite{Fedosov_RBC_2010,Fedosov_SCG_2010} 
are imposed to maintain geometric characteristics of the parasite body 
\begin{align}\label{eq:av}
U_{A,V} = \frac{k_{A,glob}(A-A_0^{tot})^2}{2A_0^{tot}}+ \nonumber \\
\sum_{m\in 1...N_t}{\frac{k_{A,loc}(A_m-A^m_0)^2}{2A^m_0}}+
\frac{k_V(V-V_0^{tot})^2}{2V_0^{tot}}, 
\end{align}
where $A$ denotes the instantaneous membrane area, $A_0^{tot}$ the target global area, $A_m$ is the area of the $m$-th triangle, $A^m_0$ 
the target area of the $m$-th triangle, $V$ the instantaneous body volume, and $V_0^{tot}$ the target volume. The coefficients $k_{A,glob}$, 
$k_{A,loc}$, and $k_V$ represent the global area, local area, and volume constraint coefficients, respectively. $N_t$ denotes the number of 
triangles within the triangulated surface.

The flagellum consists of $76$ segments containing four particles positioned at the corners of a square, and consecutive segments form four 
parallel filaments \cite{Rode_SMC_2019}. The segments are interlinked by harmonic springs with a potential given by Eq.~(\ref{eq:bond}).
The flagellum comprises a free portion at the anterior end, and a portion connected to the body, where 
two opposing filaments are embedded into the membrane network. The part attached to the body starts with a short straight segment at the  
body's posterior region and then wraps around the body, completing a half rotation before running along the radially symmetric body's 
centerline toward the anterior end with a free part extending beyond the body. The unactuated length of a trypanosome is $L_{tryp}=24\,\mu m$, 
which is used as a characteristic length scale. Two filaments of the flagellum that are incorporated into the membrane supply 
space- and time-dependent actuation wave by modifying the equilibrium length $l_0^i(s,t)$ of the harmonic springs with a phase shift 
$\phi = \pi$ between filaments
\begin{equation}
l_0^i(s,t) = \hat{l}_0^i + a_b \sin\left(2\pi \left( \frac{s}{\lambda_{in}} - f t \right) + \phi_0 \right),
\end{equation}
where $s$ is a curve-linear coordinate along the flagellum, $a_b$ is the actuation wave amplitude, $\lambda_{in}$ is the wavelength of 
the actuation wave, $f$ is the wave frequency, and $\phi_0$ is a phase shift for each actuating filament ($\phi_0=0$ for one filament and 
$\phi_0=\pi$ for the other). This mechanism generates a bending wave along the flagellum, where $\lambda_{in}$ determines the number 
of simultaneously present waves $N_{waves} = L_{flag}/\lambda_{in}$. The wave frequency $f$ defines the beating frequency, as well as 
the beat period $\tau = 1/f$, which is employed as a characteristic timescale. The beating plane is tangential to the body surface, 
since actuated filaments are embedded into the body surface. The flagellar bending wave also induces deformation of 
the cell body, which is more pronounced at the anterior region, where the body is thinner than at the posterior region.

Table~\ref{table:trypanosome} presents trypanosome model parameters both in simulation and physical units. 
\begin{table*}[t]
    \centering
    \begin{tabular}{|c|c|c|}
        \hline
        \textbf{Trypanosome parameters} & \textbf{Simulation units} & \textbf{Physical units} \\
        \hline
        $N_{tryp}$ & $1799$ &  \\
        $L_{tryp}$ & $30$ & $24 \, \mu m$ \\
        $f$ & $0.025$ & $20\, Hz$ \\
        $k_B T$ & $0.44$ & $4.28 \times 10^{-21}\, J$ \\
        $s_0$ & $1.25 \times 10^{-2}\, L_{tryp}$ &  $0.3 \, \mu m$ \\
        $R_{max}$ & $6.25 \times 10^{-2}\, L_{tryp}$ & $1.5 \, \mu m$ \\
        $K$ & $1.41 \times 10^4\, k_BTL_{tryp}$ & $1.46\, nN\mu m^2$ \\
        $\mu_{b}$ & $1.48 \times 10^7\, k_BT/L_{tryp}^2$ & $110.85\, \mu N/m$ \\
        $k_{A,glob}$ & $8.52 \times 10^6\, k_BT/L_{tryp}^2$ & $64 \times 10^{-6} \, N/m$ \\
        $k_{A,loc}$ & $8.52 \times 10^6\, k_BT/L_{tryp}^2$ & $64 \times 10^{-6} \, N/m$ \\
        $k_V$ & $2.56 \times 10^8\, k_BT/L_{tryp}^3$ & $80 \, N/m^2$\\
        $\kappa$ & $4,73 \times 10^2\, k_BT$ & $2.05 \times 10^{-18} \, J$  \\
        $a_b$ & $1.33 \times 10^{-3}\, L_{tryp}$ &   \\
        $\lambda_{in}$ & $5.33 \times 10^{-1}\, L_{tryp}$ & $12.8\, \mu m$  \\
        \hline
    \end{tabular}
    \caption{Trypanosome parameters in units of the trypanosome length $L_{tryp}$ and the thermal energy $k_BT$ with the corresponding 
    physical values. $N_{tryp}$ is the number of particles discretizing the trypanosome, $f$ is the beating frequency, 
    $s_0$ is the distance between two cross-sectional segments of the flagellum, $R_{max}$ is the maximum radius of the body, 
    $K$ is the bending rigidity of the flagellum, $\mu_{b}$ is the shear modulus of the body,  
    $k_{A,glob}$, $k_{A,loc}$, and $k_{V}$ are the local area, global area, and volume constraint coefficients, $\kappa$ is the 
    bending rigidity of the body, $a_b$ is the actuation amplitude, and $\lambda_{in}$ is the wave length. In simulations, 
    we have selected $L_{tryp}=30$, $k_BT=0.44$, and $f = 0.025$.}
    \label{table:trypanosome}
\end{table*}

\section{Red blood cell model}
\label{app:rbc}

RBCs are modelled as triangulated networks of vertices representing the cell membrane, which interact through the potential 
\cite{Fedosov_RBC_2010,Fedosov_SCG_2010}
\begin{equation}
    U_{RBC} = U_{bond,RBC} + U_{bend,RBC} + U_{A,V}.
\end{equation}
Cell in-plane elasticity is introduced through a non-linear bond potential $U_{bond,RBC} = U_{WLC} + U_{POW}$ that combines 
a worm-like-chain potential $U_{WLC}$ with a power potential $U_{POW}$ along vertex connections as
\begin{align}
    U_{WLC} &= \frac{k_BT l_m}{4p}\frac{3x^2-2x^3}{1-x}, \\
    U_{POW} &= \frac{k_p}{l},  
\end{align}
where $x=l/l_m \in (0,1)$, $l$ denotes the spring length, $l_m$ the maximum spring extension, $p$ the persistence length, 
$k_BT$ the energy unit, and $k_p$ the force coefficient. The model parameters are determined by the shear modulus which 
relates to the Young's modulus through $\mu_{RBC} = Y/4$ \cite{Dao_MBA_2006}.

The bending rigidity of the membrane is incorporated through a bending energy between adjacent triangles sharing a common edge
\begin{equation}
  U_{bend,RBC} = \sum_{j\in1..N_s}k_b[1-\cos(\theta_j-\theta_j^0)],
\end{equation}
where $k_b$ is the bending coefficient related to the bending rigidity $\kappa_{RBC} = \sqrt{3} k_b/2$, $N_s$ is the total number 
of edges, $\theta_j$ is the instantaneous angle between two adjacent triangles within a dihedral, and $\theta_j^0$ is the spontaneous 
angle between the two faces representing spontaneous curvature. Surface area and volume conservation potentials $U_{A,V}$ follow 
the same implementation as those for the parasite body in Eq.~(\ref{eq:av}). Table~\ref{table:RBC} lists RBC model parameters both 
in simulation and physical units. 

\begin{table}[t]
    \centering
    \begin{tabular}{|c|c|c|}
        \hline
        \textbf{RBC parameters} & \textbf{Simulation units} & \textbf{Physical units} \\
        \hline
        $N_{RBC}$ & $1000$ & \\
        $Y_{RBC}$ & $2.52 \times 10^{6}\, k_BT/L_{tryp}^2$ & $18.9 \, \mu N/m$ \\
        $k_{A,glob,RBC}$ & $2.8 \times 10^{7}\, k_BT/L_{tryp}^2$ & $2.1 \times 10^{-4} N/m$ \\
        $k_{A,loc,RBC}$ & $5.73 \times 10^{5}\, k_BT/L_{tryp}^2$ & $4.3 \times 10^{-6} N/m$ \\
        $k_{V,RBC}$ & $7.03 \times 10^8\, k_BT/L_{tryp}^3$ & $220\, N/m^2$  \\
        $\kappa_{RBC}$ & $69.3\, k_BT$ & $3\cdot 10^{-19}\, J$ \\
        $A_{RBC}$ & $2.32 \times 10^{-1}\, L_{tryp}^2$ & $133\, \mu m^2$ \\
        $V_{RBC}$ & $6.75 \times 10^{-3}\, L_{tryp}^3$ & $93.3\, \mu m^3$ \\
        \hline
    \end{tabular}
    \caption{RBC parameters in units of the trypanosome length $L_{tryp}$ and the thermal energy $k_BT$ with the corresponding 
    physical values. $N_{RBC}$ is the number of particles representing the RBC surface, $Y_{RBC}$ is the Youngs modulus, 
    $k_{A,glob, RBC}$, $k_{A,loc,RBC}$, and $k_{V,RBC}$ are the local area, global area, and volume constraint coefficients, 
    $\kappa_{RBC}$ is the bending rigidity of the cell, and $A_{RBC}$ and $V_{RBC}$ are the RBC surface area and cell volume.}
    \label{table:RBC}
\end{table}

\section{Fluid model}
\label{app:fluid}

Fluid is modeled using the smoothed dissipative particle dynamics (SDPD) method \cite{Espanol_SDPD_2003, Mueller_SDPD_2015}, 
where it is represented by a collection of interacting particles, each corresponding to a finite fluid volume. Interactions between particles
include conservative, dissipative, and stochastic forces. The conservative component controls fluid compressibility by enforcing
an equation of state of the form $p = p_0 (n / n_0)^\alpha - p_l$, where $n$ is the local particle number density, $n_0$ is a
reference number density, and $p_0$, $\alpha$, and $p_l$ are parameters that determine the fluid’s compressibility and set the speed of sound
via $c^2 = p_0 \alpha / n_0$. The SDPD fluid parameters are given in Table~\ref{table:fluid}. Fluid viscosity is controlled by 
the dissipative interactions, while the balance between dissipative and random forces maintains the system at a fixed temperature, 
effectively acting as a thermostat. Note that, we adopt an SDPD formulation that conserves angular momentum, in addition to mass and 
linear momentum, as described in Ref.~\cite{Mueller_SDPD_2015}. Both RBCs and the trypanosome are embedded into the SDPD fluid, such that 
particles representing the embedded elastic structures serve also as fluid discretization points.   

\begin{table}[t]
    \centering
    \begin{tabular}{|c|c|}
        \hline
        \textbf{Fluid parameters} & \textbf{Simulation units} \\
        \hline
        $\eta$ & $7,67 \times 10^4\, k_BT/(L_{tryp}^3f)$ \\
        $n_0$ & $5.4 \times 10^5\, / L_{tryp}^3$ \\
        $p_0$ & $6.14 \times 10^7\, k_B T / L_{tryp}^3$ \\
        $p_l$ & $6.01 \times 10^7\, k_B T / L_{tryp}^3$ \\
        $\alpha$ & $7$ \\
        $r_c$ & $3.21 \times 10^{-2}\, L_{tryp}$ \\
        $m_f$ & $639 \times 10^{-1}\, k_B T / (L_{tryp} f)^2$ \\
        $m_{tryp}$ & $20 \ m_f$ \\
        \hline
    \end{tabular}
    \caption{Fluid parameters of the SDPD model in units of the trypanosome length $L_{tryp}$, the thermal energy $k_BT$, and 
    the beating frequency $f$. $\eta$ is the fluid viscosity, $n_0$ is the fluid particle density, $p_0$, $p_l$ and $\alpha$ 
    control fluid compressibility, $r_c$ is the cutoff radius, $m_f$ is the mass of a fluid particle, and $m_{tryp}$ 
    is the mass of a trypanosome particle.}
    \label{table:fluid}
\end{table}

\section{Description of movies}

\begin{itemize}
    \item {\bf Movie S1}. Illustration of forward trypanosome motion through a suspension of RBCs at $H_t=0.1$. Trypanosome body is drawn in blue and the flagellum 
    in green and orange, while RBCs are depicted in red. 

    \item {\bf Movie S2}. Trypanosome propulsion through a suspension of small spheres with a radius of $R_{sph}/R_{RBC} = 0.2$ ($R_{sph} = 0.8\, \mu m$) at $\Phi = 0.1$.  
    The color code for trypanosome is the same as in Movie S1, while the spheres are drawn in red. 

    \item {\bf Movie S3}. Trypanosome propulsion through a suspension of large spheres with a radius of $R_{sph}/R_{RBC} = 1.6$ ($R_{sph} = 6.4\, \mu m$) at $\Phi = 0.1$.  
    The color code for trypanosome is the same as in Movie S1, while the spheres are drawn in red. 
    
    \item {\bf Movie S4}. Motion of a trypanosome in a slit-like environment recorded in experiments with $5\, \mu m$ beads. The video nicely shows an acceleration of the 
    trypanosome from approximately $11\,\mu m/s$ to $36\,\mu m/s$, as it enters a dense region of suspended beads. 
\end{itemize}

%

\end{document}